\documentclass[prl,aps,reprint,showpacs,amssymb,superscriptaddress]{revtex4-1}
\usepackage{graphicx}
\usepackage{dcolumn}
\usepackage{bm}
\usepackage{hyperref}
\usepackage[usenames,dvipsnames]{color}
\usepackage{amssymb}
\usepackage{amsmath}
\usepackage{amsfonts}
\usepackage{latexsym}

\newcommand{\bk}{\mathbf{k}}
\newcommand{\ua}{\uparrow}
\newcommand{\da}{\downarrow}
\newcommand{\eq}{\begin{equation}}
\newcommand{\eqx}{\end{equation}}
\newcommand{\eqn}{\begin{eqnarray}}
\newcommand{\eqnx}{\end{eqnarray}}
\newcommand{\s}{\sigma}
\newcommand{\veck}{{\bf k}}
\newcommand{\veci}{{\bf i}}
\newcommand{\vecj}{{\bf j}}

\newcommand{\vecl}{{\bf l}}
\newcommand{\la}{{\langle}}
\newcommand{\ra}{{\rangle}}

\newcommand{\g}[1]{{\bf #1}}

\begin{document}

\title{Coexistence of Nematic Order and Superconductivity in the Hubbard Model}

\author{Jan Kaczmarczyk}
\email{jkaczmar@ist.ac.at}
\affiliation{Marian Smoluchowski Institute of Physics$,$ Jagiellonian University$,$ 
ulica \L{}ojasiewicza 11$,$ PL-30-348 Krak\'ow$,$ Poland}
\affiliation{Institute of Science and Technology Austria, Am Campus 1, 3400 Klosterneuburg}

\author{Tobias Schickling}
\affiliation{Fachbereich Physik, Philipps Universit\"at Marburg,
D-35032 Marburg, Germany}

\author{J\"org B\"unemann}
\email{buenemann@gmail.com}
\affiliation{Fachbereich Physik, Philipps Universit\"at Marburg, D-35032 Marburg, Germany}

\date{\today}

\begin{abstract}
We study the interplay of nematic and superconducting order in the two-dimensional Hubbard model and show that they can coexist, especially when superconductivity is not the energetically  dominant phase. Due to a breaking of the $C_4$ symmetry, the coexisting phase inherently contains admixture of the $s$-wave pairing components. As a result, the superconducting gap exhibits very non-standard features including  changed nodal directions. Our results also show that in the optimally doped regime the superconducting phase is typically unstable towards developing nematicity (breaking of the $C_4$ symmetry). This has implications for the cuprate high-$T_c$ superconductors, 
for which in this regime the so-called intertwined orders have recently been observed. Namely, the coexisting phase may be viewed as a precursor to such more involved patterns of symmetry breaking.
\end{abstract}




\maketitle

Electronic nematic instabilities are observed across several families of correlated electron systems~\cite{Fradkin_Rev} including the cuprates~\cite{Keimer2015,2015arXiv150803075P,RevModPhys.87.457,Vojta2009}, pnictides~\cite{Chu13082010,Fernandes2014}, Ruthanate Sr$_3$Ru$_2$O$_7$~\cite{Borzi12012007}, heavy fermionic URu$_2$Si$_2$~\cite{Okazaki28012011}, as well as for dipolar gases in optical lattices~\cite{Aikawa19092014}. In the nematic (N) phase the discrete lattice rotational symmetry is lowered while the system retains  its translational symmetry. Such phases often appear close to superconductivity (SC) in the phase diagrams of these systems~\cite{Keimer2015}. A natural question therefore arises about the interplay of these two types of symmetry breakings, especially 
 since it is widely assumed that both can have a common cause, a large on-site Coulomb interaction~\cite{Yokoyama2,Buenemann2012}.
 This question is especially relevant for the high-$T_c$ cuprates, as in the optimal-doping regime both orderings were reported~\cite{Keimer2015} (in different parameter ranges),
%
%
as were also other microscopic orders. The coexistence of a number of them has led to the notion of `intertwined orders'~\cite{Keimer2015,RevModPhys.87.457}. Here we demonstrate the coexistence of the SC and N orderings in the minimal model for the description of strongly correlated/high-$T_c$ systems, the single-band Hubbard model. 
  
Superconductivity and nematic order (in the form of an electronic `Pomeranchuk instability'~\cite{Pomeranchuk,PhysRevLett.85.5162, Yamase2000a, Yamase2000b, PhysRevB.81.073108}) 
have frequently been considered as competing phenomena~\cite{Yamase2000b,PhysRevB.81.073108, PhysRevB.73.174513,PhysRevB.74.165109}. Consequently, the coexisting (N+SC) phase was studied usually for anisotropic models, where its description is much easier, as the rotational symmetry is broken already in the starting Hamiltonian~\cite{Yamase2000b,PhysRevB.73.174513,PhysRevB.73.214517,PhysRevB.74.165109, PhysRevB.82.180511,PhysRevB.84.220506,PhysRevB.91.195121}. On the other hand, 
for isotropic models a N+SC phase was studied only in rather specific situations using (i) the perturbation expansion method~\cite{PhysRevB.67.035112}, applicable only in the weak-coupling limit, and (ii) a phenomenological model where interactions leading to both orderings were postulated separately~\cite{PhysRevB.75.155117}. 

The current state-of-the-art methods for strongly correlated electron systems are generally not well-suited to capture the subtle effects of the \textit{correlation-induced} Fermi surface deformations related to a N phase, let alone those of a N+SC phase. The limitations stem mostly from finite system sizes, which translate to a low momentum  space ($\bk$-space) resolution \footnote{cf. e.g. Fig.~2(b) of Ref.~\cite{PhysRevB.74.165109}}. For example, in Dynamical-Mean-Field Theory~\cite{PhysRevB.82.180511} and Dynamic Cluster Approximation~\cite{PhysRevB.82.180511,PhysRevB.84.220506,PhysRevB.80.245102} the system size is up to $4 \times 4$
~\footnote{These methods suffer also from the sign problem, which can significantly restrict the temperature and parameter range and, as a result, can render the analysis of stabilities of considered phases difficult or impossible (cf. the discussion in Refs.~\cite{PhysRevB.84.220506,PhysRevB.80.245102})}
and in variational Monte Carlo (VMC)~\cite{PhysRevB.74.165109, PhysRevB.92.085109, Zou2015} up to $24 \times 24$. Consequently, finite-size errors can be significant even for anisotropic models \footnote{cf. e.g. Fig.~7 of Ref.~\cite{PhysRevB.74.165109} or Fig.~1 of Ref.~\cite{PhysRevB.92.085109}} and may even lead to qualitatively different results depending on the system size~\cite{PhysRevB.84.220506}.
Hence, it is an entirely open question in correlated-electron theory whether the SC and the N phase are generically competing or tend to stabilize each other~\cite{PhysRevB.85.184511, Liu2013, PhysRevLett.114.097001, PhysRevB.88.180502}.

  In this work, we overcome the difficulties in describing a N+SC phase and show that N and SC orderings can coexist in the Hubbard model. To this end, we use a variational method  based on Gutzwiller wave functions (GWF)~\cite{Gutzwiller} combined with an efficient {\it diagrammatic expansion} (DE) technique~\cite{Florian,Buenemann2012}, which enables one to evaluate expectation values for GWF without any additional uncontrolled approximations.
  This DE-GWF method has been applied  successfully to study Fermi-surface deformations, $d$-wave superconductivity, and quasiparticle band structures in the Hubbard~\cite{Buenemann2012,Kaczmarczyk2013,Kaczmarczyk2015a,Kaczmarczyk2015b}, $t$-$J$~\cite{Kaczmarczyk2014}, Anderson lattice~\cite{PhysRevB.92.125135, 2015arXiv151000224W}, and multiband~\cite{2015arXiv151007896M} models. The method works in the thermodynamic limit, i.e., with {\it no finite size limitations}, which enables us to properly investigate the stability of a N+SC phase. 
In the single-band Hubbard model the dominant nematic order has a $d$-wave form~\cite{Buenemann2012, PhysRevLett.85.5162}, 
meaning that the Fermi surfaces are stretched along one lattice axis and compressed along the other. Combined with a $d$-wave superconducting pairing,  symmetry requires that there is an additional induced $s$-wave component of the SC gap with both on-site and long-range contributions (the on-site contribution is often neglected~\cite{Yamase2000b,PhysRevB.74.165109}). This leads to a non-trivial gap structure with features such as shifted nodal points and modified zero-gap regions, see below.

Our starting point is the Hubbard model on a two-dimensional, infinite square lattice, as given by the Hamiltonian
\eq
\hat{H}=\sum_{\veci,\vecj,\sigma}t_{\veci,\vecj}\hat{c}_{\veci,\sigma}^{\dagger}
\hat{c}_{\vecj,\sigma}^{\phantom{\dagger}} + U\sum_{\veci} \hat{d}_{\veci}
\,\, , \quad
 \hat{d}_{\veci}\equiv\hat{n}_{\veci,\uparrow}\hat{n}_{\veci,\downarrow}\, ,
 \label{eq:HM}
\eqx
where $\veci = (i_x, i_y)$ is the two-dimensional site-index, $t_{\veci \vecj} = -t$ and $t'$ are the hopping integrals for the nearest and next-nearest neighbors, respectively, $U$ is the Coulomb interaction, and $\s = \ua, \da$ is the spin quantum number.

To account for electronic correlations, the strength of which is determined by the ratio of $U/|t|$, a `Jastrow correlator' is used $|\Psi_{\rm G}\rangle=\hat{P}_{\rm G}|\Psi_0\rangle$, where $|\Psi_0\rangle$ is a single-particle-product wave function (Slater determinant) to be defined later. 
We work with the Gutzwiller correlator $\hat{P}_{\rm G} = \prod_{\veci}\hat{P}_{\veci}$, in which the local correlators can be expressed as $\hat{P}_{\veci}=\sum_{\Gamma}\lambda_{\Gamma}
|\Gamma \rangle_{\veci\,\veci}\! \langle \Gamma |
+\lambda_B (|d \rangle_{\veci\,\veci}\! \langle \emptyset |+{\rm H.c.})$. 
The parameters $\lambda_\Gamma$ control occupancies of the four local states $|\Gamma \rangle_{\veci}$, whereas $\lambda_B$ is related to the on-site pairing component~\cite{Buenemann2005}.
The principal task is the evaluation of the expectation value $E_{\rm G}$ of the Hamiltonian 
 $\hat{H}$ with respect to the Gutzwiller wave function $|\Psi_{\rm G}\rangle$. 
This evaluation remains a difficult many-particle problem. 
It has been shown in~\cite{Florian,Buenemann2012} that an efficient {\it diagrammatic expansion} scheme can be formulated for this purpose if the local correlator is chosen such that it fulfils the condition
\eq
\hat{P}_{\veci}^{\dagger} \hat{P}_{\veci}^{}
=\hat{P}^2_{\veci}=1+x\,\hat{d}_{\veci}^{\rm HF} \; ,
\label{eq:1.4}
\eqx
where $x \in [-4, 0]$ is a variational parameter and the Hartree-Fock (HF) operators are defined by
\eq
\hat{d}^{\rm HF}_{\veci} = \hat{d}_{\veci}-n_0(\hat{n}_{\veci,\uparrow}+\hat{n}_{\veci,\downarrow})
-\Delta_0(\hat{\Delta}_{\veci}+\hat{\Delta}^{\dagger}_{\veci})+
n^2_0+\Delta^2_0 \, .
\eqx
Here $n_0 \equiv \la \hat{n}_{\veci,\sigma} \ra_0$ and we already allow for a breaking of the $C_4$ symmetry which, as mentioned above, leads to a finite on-site pairing $\Delta_0 \equiv \la \hat{\Delta}_{\veci} \ra_0 = \la \hat{c}_{\veci, \da} \hat{c}_{\veci, \ua} \ra_0 $~\footnote{{It also leads to non-zero double occupancies, $\langle \hat{d}_G \rangle \neq 0$, and makes the analysis of a N+SC phase within the $t$-$J$ model~\cite{Yamase2000b,PhysRevB.74.165109} inconsistent with the zero double occupancy condition.}}. In the following, we use the notation $\langle \ldots \rangle_{0,{\rm G}}$ for expectation values with respect to $|\Psi_0\rangle $ and $|\Psi_{\rm G}\rangle$. With our choice of a correlator which satisfies~(\ref{eq:1.4}) we eliminate on-site terms (the so-called `Hartree bubbles') from the resulting diagrammatic expansion of expectation values. As a consequence, the results of the DE-GWF method converge rapidly 
 with an increasing order of the expansion parameter $x$, as was demonstrated for one-dimensional systems~\cite{Buenemann2012}. 
 Note that with the condition~(\ref{eq:1.4}), the parameters $\lambda_{\Gamma}$, $\lambda_{B}$ in our Gutzwiller correlator are (for a given $| \Psi_0 \rangle$)
 all determined as a function of $x$, which serves as our only remaining variational parameter.  
The DE-GWF method is systematic in the sense that in the zeroth order of the expansion it reproduces~\cite{Kaczmarczyk2014,PhysRevB.92.125135} the non-trivial results of the Gutzwiller approximation whereas, with an increasing order, the exact GWF solution is approached. For two-dimensional systems DE-GWF 
gives results in agreement with VMC but with better accuracy~\cite{Kaczmarczyk2013,Kaczmarczyk2014}.

Within the DE-GWF method we obtain all expectation values, for example $E_{\rm G}$, as a power series in $x$, 
\begin{equation}
E_{\rm G}(|\Psi_0\rangle, x)\approx \sum_{k=0}^{k_{\rm c}} \tilde{e}_{k} \frac{x^{k}}{k!}\;.\label{dd}
\end{equation}
The explicit form of $E_{\rm G}$ for states with a finite onsite pairing $\Delta_0$ is given in~\cite{Kaczmarczyk2015b}. 
The coefficients $\tilde{e}_{n}$ depend on the wave function $|\Psi_0\rangle$ or, more precisely, on the expectation values
\begin{equation}
\begin{gathered}
 P^{\sigma}_{\vecl,\vecl'} \equiv  P_{\vecl,\vecl'} \equiv
\langle\hat{c}^{\dagger}_{\vecl,\sigma}\hat{c}_{\vecl',\sigma}\rangle_0\,, 
\quad  
S_{\vecl,\vecl'} \equiv 
\langle\hat{c}_{\vecl,\da}\hat{c}_{\vecl',\ua}\rangle_0\,.
 \label{eq:lines}
 \end{gathered}
 \end{equation}
The intersite expectation values serve as lines in our diagrammatic expansion.
The number of lines in the diagrams grows with the order $k$. Instead of terminating the expansion~(\ref{dd}) with some finite value of $k_{\rm c}$, it turns out to be more accurate to 
 include all diagrams up to a certain maximum number of lines $l_{\rm c}$. We further need to introduce a real-space cutoff, 
 i.e., we 
only include lines up to the maximum distance, here $|\g l -\g l '|^2\equiv (l_x-l'_x)^2+
(l_y-l'_y)^2 = 16$ (measured in lattice constants).
 
In the presence of superconductivity we minimize the functional $\mathcal{F} \equiv E_{\rm G} - 2 \mu_{\rm G} \la \hat{n}_\s \ra_{\rm G}$ instead of $E_{\rm G}$~\cite{Kaczmarczyk2013,Kaczmarczyk2014}, where $\mu_{\rm G}$ is the chemical potential. The minimization with respect to 
 $|\Psi_0\rangle $ leads to the effective single-particle equation (cf. Appendix A of Ref.~\cite{Schickling2014} and Ref.~\cite{PhysRevLett.100.016405})
\eq
\hat{H}_0^{\rm eff} |\Psi_0\ra = E |\Psi_0\ra,\label{sa1}
\eqx
where the effective Hamiltonian is given as
\eq
\hat{H}^{\rm eff}_0=
\sum_{\veci,\vecj,\sigma}t^{\rm eff}_{\veci,\vecj}
\hat{c}_{\veci,\sigma}^{\dagger}\hat{c}_{\vecj,\sigma}^{\phantom{\dagger}}
+ \sum_{\veci, \vecj} \bigl( \Delta^{\rm eff}_{\veci,\vecj}
\hat{c}_{\veci,\ua}^{\dagger}\hat{c}_{\vecj,\da}^{\dagger} +
{\rm H.c.} \bigr) \label{eq:effH}\,.
\eqx
Here we introduced the effective hopping and pairing parameters
\eq
t^{\rm eff}_{\veci,\vecj} =
\frac{\partial \mathcal{F} (|\Psi_0\rangle,x)}{\partial P_{\veci,\vecj}}\;,
\quad 
\Delta^{\rm eff}_{\veci,\vecj} =
\frac{\partial \mathcal{F} (|\Psi_0\rangle,x)}{\partial S_{\veci,\vecj}} \;.
\label{eq:eff}
\eqx
Let us underline that these parameters contain long-range components, with the same cutoff as for the lines (i.e. up to $|\veci-\vecj|^2 = 16$). Such long-range components are usually neglected in other methods, but they turn out to be important for a proper description of the nematic phases.
The remaining task is the self-consistent solution of Eqs.~(\ref{sa1})-(\ref{eq:eff}) in $\veck$-space, together with the 
 minimization condition $\partial \mathcal{F} / \partial x = 0$ (see Refs. \cite{Kaczmarczyk2015a, Kaczmarczyk2015b, Kaczmarczyk2014} for details on the numerical procedures). From the final self-consistent solution we can calculate the effective dispersion $\epsilon_\bk^{\rm eff}$ and gap $\Delta_\bk^{\rm eff}$ as Fourier transforms of $t^{\rm eff}_{\veci,\vecj}$ and~$\Delta^{\rm eff}_{\veci,\vecj}$, as well as the ground-state energy $E_{\rm G}$ labelled $E_{\rm N}$, $E_{\rm SC}$, and $E_{\rm N+SC}$ for the three considered phases.

We select the value of the nearest-neighbor hopping, $t$, as our unit of energy and choose the typical values of other parameters reflecting the cuprate high-$T_c$ superconductors, $U = 10$ and $t' = -0.25$ (unless stated otherwise). Physical energies (in Kelvins) are obtained by assuming $t = 350 \, {\rm meV}$. The diagrammatic expansion was carried out up to terms with $l_{\rm c} = 13$ lines in the diagrams.

\begin{figure}[t]
\includegraphics[width=0.96\columnwidth]{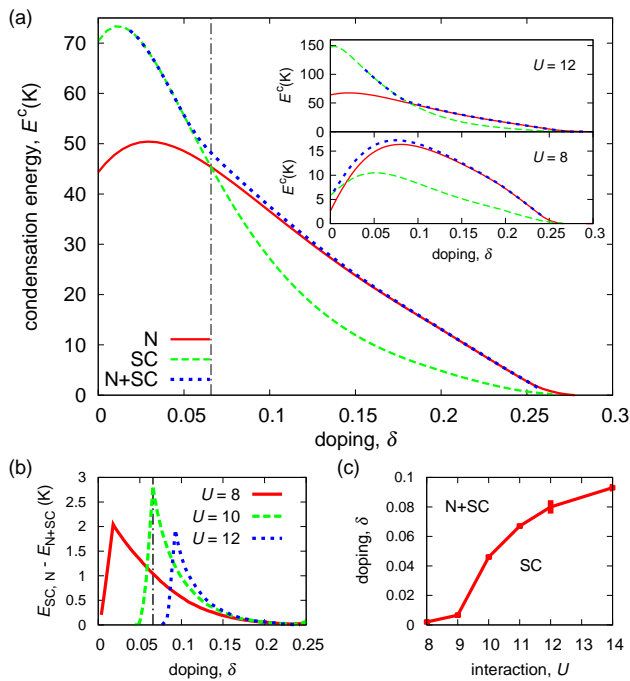}
\caption{(color online).
(a) Condensation energies of the considered phases: N - nematic, SC - superconducting, and a N+SC phase with coexisting orders. The insets show the same quantities for $U=12$ and $U=8$. (b) The energy gain from developing the second order on top of the optimal one-order phase (N or SC): ${\rm min}(E_{\rm SC}, E_{\rm N}) - E_{\rm N+SC}$. (c) Phase diagram as a function of doping and the Coulomb interaction.
\label{fig1}}
\end{figure}

In Fig.~\ref{fig1} we compare the condensation energies $E^{\rm c}_{\rm N,SC,N+SC}$ (=energy gain relative to the phase without broken symmetry) of the three studied phases.
We find the best conditions for a N+SC phase when $E^{\rm c}_{\rm N} \approx E^{\rm c}_{\rm SC}$ as, e.g., for $\delta \approx 0.07$ in Fig.~\ref{fig1}(a) (marked with dot-dashed lines in this and some of the following figures). When $E^{\rm c}_{\rm SC}$ is significantly higher than $E^{\rm c}_{\rm N}$  the coexisting phase becomes unstable, as for $\delta \lesssim 0.05$. In the regime of larger doping, where $E^{\rm c}_{\rm N} > E^{\rm c}_{\rm SC}$,
an additional SC ordering on top of the N phase is stable even when the pure SC phase has a significantly lower condensation energy, as it is the case for $\delta>0.1$. In fact, for the optimally-doped case the SC phase is higher in energy than the N (or N+SC) phase by $\sim 10-12 \, {\rm K}$ independent of the expansion cutoff (as verified for $l_{\rm c}=13,11,9$).
The energy gain from developing additional order on top of the `optimal' one-order phase (SC or N) is similar for $U=8, 10 \textrm{, and } 12$ and maximally equal to $2-3$~K in the vicinity of the crossing of the SC and N phase energies, as visualized in Fig.~\ref{fig1}(b).
Figure~\ref{fig1}(c) shows how the boundary between the SC and N+SC phases evolves with~$U$, which is closely related to the crossing of the N and SC phases [cf. Fig.~\ref{fig1}(a)]. 
This picture would be changed if the spin-exchange term was introduced into the starting Hamiltonian (i.e. in the $t$-$J$-$U$ model) to make up for the underestimation of spin-exchange effects by GWF. Such term favors the SC phase and we find that a N+SC phase is stable up to $J \approx 0.15$, above which the pure SC phase dominates. On the other hand, the N phase induces a distortion of the underlying lattice of ions~\cite{Fradkin_Rev}, which should favor the N and N+SC phases.
To account for such effects the inclusion of electron-phonon coupling is required~\cite{PhysRevB.89.195139,PhysRevB.91.205135}, which is beyond the scope of the present paper.

\begin{figure}[t!]
\includegraphics[width=0.95\columnwidth]{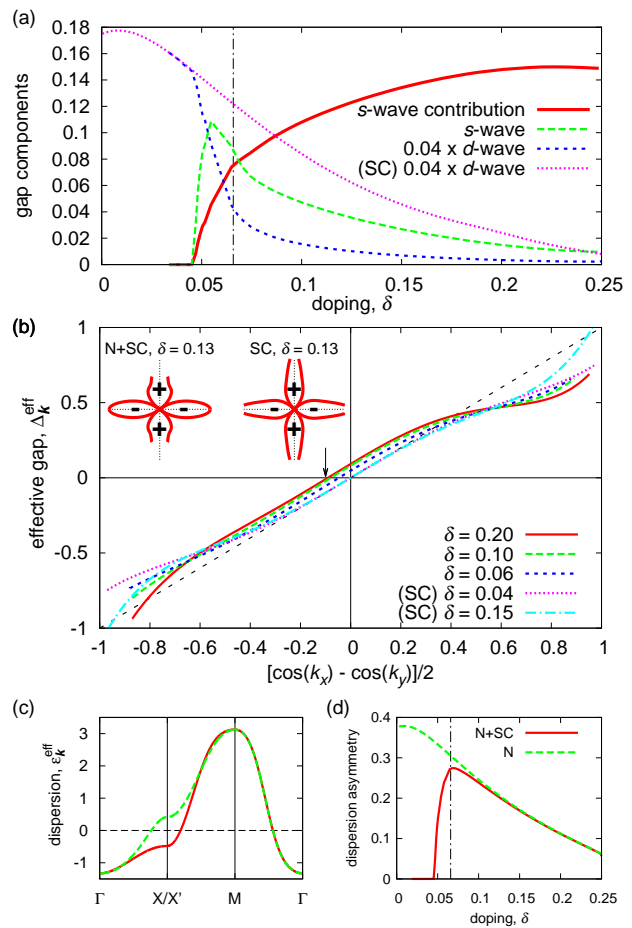}
\caption{(color online).
(a) Gap components integrated over the Fermi surface (see the main text for details). 
(b) Gap magnitude (normalized) along the Fermi surface (dashed line shows the $d_{x^2-y^2}$ gap). The shifted nodal point is marked by an arrow. Inset: polar plots of the $s+d$-wave gap of a N+SC phase and the $d$-wave gap of the SC phase, both at $\delta = 0.13$. (c) Dispersion along two paths in the Brillouin zone, with the points $\Gamma=(0,0)$, $X=(0,\pi)$, $X'=(\pi, 0)$, and $M=(\pi, \pi)$. (d) Dispersion asymmetry defined in the main text.
\label{fig2}}
\end{figure}

In Fig.~\ref{fig2} we elucidate the non-standard gap structure in a N+SC phase. Breaking of the $C_4$ symmetry induces an additional $s$-wave component of the SC gap. In Fig.~\ref{fig2}(a) we show the integral of the magnitude of the $s$- and $d$-wave gap components, defined as $|\Delta^{\rm eff}_{(k_x, k_y)} \pm \Delta^{\rm eff}_{(k_y, k_x)}|$ along the Fermi surface. 
The `$s$-wave contribution' curve shows the ratio of the $s$-wave integral to the sum of the two integrals to quantify the $s$-wave input to the pairing. It is equal to $0$ ($1$) for a pure $d$($s$)-wave state. Strikingly, although the energy gain from developing a SC gap on top of the N phase is rather small (maximally $\sim 3$ K), the value of the effective gap can be of the same order of magnitude as that in the pure SC phase. 
The deviation of the gap from the standard $d_{x^2-y^2}$ behavior along the Fermi surface is shown in Fig.~\ref{fig2}(b). Such deviation is present even for the SC phase~\cite{Kaczmarczyk2013,Kaczmarczyk2014}, as also observed experimentally~\cite{Mesot1999,Kondo2009,Hashimoto2014}. For a N+SC phase additionally the nodal point is slightly shifted away from the diagonal direction (as marked with an arrow), in contrast to previous results~\cite{PhysRevB.84.220506} for an anisotropic model, and the effect increases with doping. The inset shows the polar plots of the gap for N+SC and SC phases. In the former case the Fermi surface is open in the vertical and closed in the horizontal direction.
To quantify the nematicity of the system we plot in Fig.~\ref{fig2}(c) the dispersion relation along high-symmetry lines 
for $\delta=0.13$. For the case without nematicity, the dispersions at the $X$ and $X'$ points would be equal. The difference in the values of dispersion at these two points divided by the bandwidth is shown in Fig.~\ref{fig2}(d) as a measure of the dispersion asymmetry for N+SC and N phases. The SC order does not modify the dispersion asymmetry significantly (nor the Fermi surface), unless the SC phase has lower energy than the N phase.

\begin{figure}[t!] 
\includegraphics[width=\columnwidth]{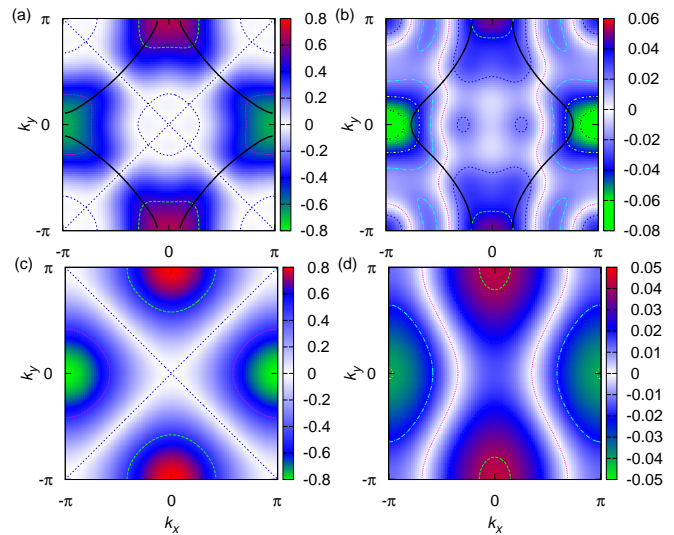}
\caption{(color online). Gap value in the Brillouin zone
(a)-(b) from the DE-GWF method for the SC (a) and N+SC (b) phases at the doping of $\delta=0.04$ and $\delta=0.13$, respectively; (c)-(d) for the gap with contributions up to nearest neighbors, namely a pure $d_{x^2-y^2}$ gap (c), and a gap with the on-site component and the nearest-neighbor $s$+$d$-wave components (d). The solid black lines show the Fermi surface in (a)-(b). The thin dashed/dotted lines are isolines and show, in particular, the regions with zero gap (white color in all graphs).
\label{fig3}}
\end{figure}

In Fig.~\ref{fig3} we show the effective gap in the Brillouin zone obtained from the DE-GWF method (with long-range contributions up to~$\Delta^{\rm eff}_{4,0}$) for the SC phase in Fig.~\ref{fig3}(a) and N+SC phase in Fig.~\ref{fig3}(b), as well as the corresponding gap structures with contributions only up to nearest neighbors for both phases in Fig.~\ref{fig3}(c)-(d), as usually assumed in other methods (e.g. in VMC). 
It can be seen from Fig.~\ref{fig3} that such an assumption does not reflect all principal features of the optimal variational solutions. For example, the longer-range components of the gap are mostly opposite to the dominant $\Delta^{\rm eff}_{1,0}$ component and this leads to circles with zero gap around the $\Gamma=(0,0)$ and $M= (\pi, \pi)$ points of the Brillouin zone in Fig.~\ref{fig3}(a). For a N+SC phase the gap structure is significantly modified with respect to the pure SC phase: (i) the magnitudes of the gap values at $X = (0, \pi)$ and $X' = (\pi, 0)$ are different; (ii) the zero-gap direction is no longer a straight line along the diagonal but an irregular line along one of the axes (coinciding with the direction, in which the Fermi surface is open); (iii) 
a larger part of the Brillouin zone contributes to pairing.

The coexistence of the nematic and superconducting orders has been demonstrated in the Hubbard model by using the full Gutzwiller wave function (GWF). Application of the diagrammatic-expansion (DE) technique has enabled us to investigate the properties of the system without finite-size limitations, a condition crucial for the description of the nematic phases. We have shown that the superconducting and nematic orders coexist in the Hubbard model unless the pure superconducting phase is significantly lower in energy than the pure nematic phase. We have obtained the phase diagram, the energies and other properties of the investigated pure and coexisting phases. The gap structure in the coexisting phase is unconventional due to the breaking of the $C_4$ symmetry: the induced $s$-wave gap component shifts the nodal point away from the diagonal direction and modifies the zero-gap region to form in the direction of open Fermi~surface. 

In the optimal doping regime pure superconductivity turns out not to be the dominating phenomenon as it is unstable against a $d$-wave nematic instability with (as well as without) an additional superconducting order. This observation may be related to the fact that the cuprate high-$T_c$ superconductors develop additional orders in this regime. Namely, the investigated phase can be viewed as a precursor to more complicated orders including stripes and phases with charge density wave order involving more complex patterns of symmetry breaking. 

The developed formalism can also be applied to other situations including dipolar Fermi gases in optical lattices where the anisotropy of dipolar interactions leads to appearance of the N phase~\cite{Aikawa19092014} and superconductivity can be induced by an attractive on-site interaction~$U$.

{\sl Acknowledgements.}
The authors are grateful to Florian Gebhard and Mikhail Lemeshko for discussions and critical reading of the manuscript.
The work was supported by the Ministry of Science and Higher Education in
 Poland through the Iuventus Plus grant No. IP2012 017172 for the years 
2013-2015, as well as by the People Programme (Marie Curie Actions) of the European Union's Seventh Framework Programme (FP7/2007-2013) under REA grant agreement n$^{\rm o}$ [291734]. JK acknowledges hospitality of the Leibniz Universit\"{a}t in Hannover where a large part of the work was performed.

\bibliography{POMSC} 

\end{document}